\documentclass[12pt]{iopart}

\usepackage{iopams}

\usepackage{bm}
\usepackage{dcolumn}
\usepackage{graphicx}

\begin{document}

\title{Electric polarization of magnetic domain walls in magnetoelectrics}

\author{I~P~Lobzenko, P~P~Goncharov and N~V~Ter-Oganessian}
\address{Institute of Physics, Southern Federal University, 194 Stachki pr., Rostov-on-Don, 344090 Russia}
\ead{nikita.teroganessian@gmail.com}

\begin{abstract}
Two prominent magnetoelectrics MnWO$_4$ and CuO possess low-temperature commensurate paraelectric magnetically ordered phase. Here using Monte Carlo simulations we show that the walls between the domains of this phase are ferroelectric with the same electric polarization direction and value as those in the magnetoelectric phases of these compounds. We also suggest that experimental observation of electric polarization of domain walls in MnWO$_4$ should help to determine the macroscopic interactions responsible for its magnetoelectric properties.
\end{abstract}


\maketitle

\section{Introduction}

Increasing demand in miniaturization of functional elements of devices has tremendously stimulated research in the physics of domain walls in recent decades~\cite{Catalan_Seidel_Review}. The walls separating different domains are often found to have new properties different from those of the domains. The examples of such exploitable phenomena are ferroelectric domain walls separating paraelectric domains~\cite{Tagantsev_Nature_Comm,Salje_SrTiO3,VanAert_CaTiO3,Picozzi_CaTiO3,Wei_PbZrO3}, conducting domain walls between insulating domains~\cite{Seidel_Wall_Conduction}, and superconducting twin walls~\cite{Salje_Review}. Thus, domain walls provide new physics, which can be used to create new types of devices, such as, for example, the racetrack memory~\cite{Parkin_RaceTrack} and magnetic domain wall logic~\cite{Allwood_Logic} or can be used as injectors/detectors of spin currents~\cite{Savero_Torres_DW_injector_detector}.

At the same time the last decade has seen rapid increase of interest in materials showing multiferroic properties~\cite{Khomskii_Review}. Ferroelectromagnets (i.e. materials showing independent magnetic and ferroelectric orders) and magnetoelectrics (i.e. materials displaying magnetically induced ferroelectric order) are of particular interest due to the possibility of creating new types of devices such as, for example, devices for reading, writing, and storage of information~\cite{Wang_Memory}, various sensors~\cite{Chen_Sensor}, microwave~\cite{Pyatakov_UFN_Review} and spintronics~\cite{Pantel_Spintronics} devices, and wireless energy transfer and energy harvesting technologies~\cite{Pyatakov_UFN_Review}.

Technological applications of multiferroics discovered to date, however, are still significantly hindered by a number of their properties. Ferroelectromagnets usually show high temperatures of ferroelectric and magnetic phase transitions and high electric polarization. However, the usually large difference between the temperatures of the magnetic and ferroelectric phase transitions and the difference in the sources of the two orders result in low magnetoelectric response~\cite{Khomskii_Review}. In turn, magnetoelectrics show direct coupling between magnetic order and induced electric polarization, but possess low magnetic phase transition temperatures and low electric polarization values. The magnetoelectric with the highest phase transition temperature discovered to date is cupric oxide CuO~\cite{KimuraCuO}, which displays modulated magnetic order and ferroelectricity between approximately 210 and 230~K.

The microscopic nature of the magnetoelectric effect is usually interpreted using the inverse Dzyaloshinskii-Moriya interaction~\cite{SergienkoDMI} or the spin-current model~\cite{KatsuraSpinCurrent}. In both models the electric polarization arises as the result of interaction of two neighboring noncollinear spins. However, single ion contribution to the microscopic origin of the magnetically induced polarization was shown to exist~\cite{Sakhnenko_Microscopy}.

Common occurrence of modulated magnetic structures in magnetoelectrics has led to pervasive notion that incommensurate magnetic order such as, for example, spiral or cycloidal is necessary to induce electric polarization~\cite{Kimura_Review_Spiral}. However, ferroelectricity can be induced by commensurate magnetic structures, which is observed, for example, in RMn$_2$O$_5$ (R = rare-earth) magnetoelectrics~\cite{Kimura125manganites}. The intimate coupling between modulated magnetic order and magnetically induced electric polarization can be understood as the symmetry property of the relevant magnetic order parameters, which allow Lifshitz invariants responsible for modulation and induce improper ferroelectric phases~\cite{SakhnenkoImproperFerroelectric}.

Two approaches exist to macroscopically describe the magnetoelectric interaction. The first one consists in inclusion of the terms in the thermodynamic potential, which can be schematically written as
\begin{equation}
P_\alpha\eta_1\frac{\partial \eta_2}{\partial \beta},\label{eq:ME_inter_Flexo}
\end{equation}
where $\alpha$ and $\beta$ are coordinates, $P_\alpha$ is the electric polarization component, and $\eta_1$ and $\eta_2$ are either different components of a single magnetic order parameter or two different magnetic order parameters. Such magnetoelectric effect due to inhomogeneous magnetic order was first studied in~\cite{Baryakhtar}, where the calculation of electric polarization of Bloch and N\'{e}el magnetic domain walls was outlined. Subsequently, the same ideas were used to describe the bulk magnetoelectric effect in spiral magnets~\cite{Mostovoy} and led to the appearance of the term {\it flexomagnetoelectric} effect~\cite{Zvezdin_Flexo}.

The second approach consists in considering the contribution to the thermodynamic potential of the form
\begin{equation}
P_\alpha\eta_1\eta_2.\label{eq:ME_inter_Trilinear}
\end{equation}
Such macroscopic coupling was successfully used to interpret magnetoelectric phenomena in a number of magnetoelectrics~\cite{Harris_Review_Order_parameters,Toledano_MnWO4,Toledano_CuO,Ter-Oganessian_Interpretation}. Apparently both interactions~(\ref{eq:ME_inter_Flexo}) and~(\ref{eq:ME_inter_Trilinear}) can describe magnetoelectric phenomena in many magnetoelectrics. However there is an important difference between them consisting in the fact that~(\ref{eq:ME_inter_Flexo}) requires spatially varying magnetic order parameters to induce electric polarization. The experimental fact that ferroelectricity occurs mostly in spatially modulated magnetic structures together with the growing number of crystals showing polar magnetic domain walls, which is explained using flexomagnetoelectric interaction, give evidence in support of the interaction~(\ref{eq:ME_inter_Flexo}). However, the interaction~(\ref{eq:ME_inter_Trilinear}) is capable of describing magnetoelectric effect in both incommensurate and commensurate magnetic structures and, therefore, provides equal footing for such description.

The purpose of this work is twofold. Firstly, we show that the walls separating the domains of low-temperature commensurate paraelectric magnetic phases of MnWO$_4$ and CuO are polar. Secondly, we show that this fact can be used to experimentally distinguish between the cases of interactions~(\ref{eq:ME_inter_Flexo}) and~(\ref{eq:ME_inter_Trilinear}) in theoretical description of magnetoelectric properties of MnWO$_4$. The paper is organized as follows. In \sref{sec:SpinHam_and_Monte} we describe the spin Hamiltonian and the results of the Monte Carlo simulations, in \sref{sec:Discussion} we discuss the obtained results, whereas in \sref{sec:Conclusions} we present the conclusions of our work.

\section{Spin Hamiltonian and Monte Carlo studies\label{sec:SpinHam_and_Monte}}

MnWO$_4$ possesses a monoclinic structure with the monoclinic angle $\beta\approx91^\circ$ described by the space group P2/c (C$_{2h}^4$)~\cite{Lautenschlager_MnWO4}. Upon lowering the temperature it undergoes a sequence of magnetic phase transitions at $T_{\rm N}$=13.5~K, $T_2$=12.7~K, and $T_1$=7.6~K, which lead to the appearance of magnetically ordered states \textbf{AF3}, \textbf{AF2}, and \textbf{AF1}, respectively~\cite{Taniguchi06_MWO}. The structure of the low-temperature commensurate magnetic phase \textbf{AF1} is described by the wave vector $\vec{k}_{\rm c}=(\frac{1}{4},\frac{1}{2},\frac{1}{2})$, whereas the incommensurate phases \textbf{AF2} and  \textbf{AF3} by the wave vector $\vec{k}_{inc}=(-0.214,\frac{1}{2},0.457)$~\cite{Lautenschlager_MnWO4}. The phase \textbf{AF2} is ferroelectric with polarization along the $b$ axis~\cite{Taniguchi06_MWO}. In the phases \textbf{AF1} and \textbf{AF3} without applied external magnetic fields the spins in MnWO$_4$ are directed along the easy axis in the $ac$ plane making an angle of about 34$^\circ$ with the $a$ axis~\cite{Sagayama_MnWO4}. In the phase \textbf{AF2} additional spin component along $b$ appears. In the following we assume that the axes $x$, $y$, and $z$ are directed along the easy axis, $b$ axis, and perpendicular to the easy and $b$ axes, respectively.

For the analysis of the magnetic structure of domain walls in MnWO$_4$ we employ Monte Carlo simulations of the Heisenberg Hamiltonian
\begin{eqnarray}
H&=&-\frac{1}{2}\sum_{\bm{R},\bm{R'},\bm{t},\bm{t'}}J(\bm{R}+\bm{t}, \bm{R'}+\bm{t'}) \bm{S}(\bm{R}+\bm{t})\bm{S}(\bm{R'}+\bm{t'})\nonumber\\
&&+\frac{1}{2}\sum_{\bm{R},\bm{t},\alpha}D_\alpha S_\alpha^2(\bm{R}+\bm{t}),\label{eq:Hamiltonian}
\end{eqnarray}
where $\bm{S}(\bm{R}+\bm{t})$ are classical spins of length $\frac{5}{2}$, $\bm{R}$ and $\bm{R'}$ are lattice vectors, $\bm{t},\bm{t'}=\bm{t_1},\bm{t_2}$ give the positions of the two Mn$^{2+}$ ions in the unit cell, $J(\bm{R}+\bm{t}, \bm{R'}+\bm{t'})$ is the superexchange constant between the spins at $\bm{R}+\bm{t}$ and $\bm{R'}+\bm{t'}$, $D_\alpha$ is the single-ion anisotropy, and $\alpha=x,y,z$. The first sum in~(\ref{eq:Hamiltonian}) is evaluated over pairs of spins.

Inelastic neutron scattering was used to determine up to eleven superexchange constants in MnWO$_4$~\cite{Ehrenberg_MWO,Ye_MWO}, which are summarized in \tref{tab:Exchange_Constants_MnWO4}.
\begin{table}
\caption{Magnetic superexchange constants for the Mn$^{2+}$ ions in MnWO$_4$ given in units of meV. The Mn -- Mn distance is given in \AA. \label{tab:Exchange_Constants_MnWO4}}
\begin{indented}
\item[]\begin{tabular}{c ccc ccc}
\br
 & $J_1$ & $J_2$ & $J_3$ & $J_4$ & $J_5$ & $J_6$ \\
\mr
Mn -- Mn & 3.28 & 4.4 & 4.82 & 4.99 & 5.75 & 5.8 \\
\cite{Ehrenberg_MWO}&-0.168&-0.116&-0.364&0.356&0.018&-0.438\\
\cite{Ye_MWO}&-0.84&-0.08&-0.64&-0.52&0.1&-0.86\\
\mr
\mr
 & $J_7$ & $J_8$ & $J_9$ & $J_{10}$ & $J_{11}$ & $D$ \\
\mr
Mn -- Mn & 5.87 & 6.49 & 6.56 & 6.88 & 7.01 &   \\
\cite{Ehrenberg_MWO}&0.02&0.424&-1.96&&&-0.122\\
\cite{Ye_MWO}&-0.24&0.04&-0.52&-0.3&0.04&-0.18\\
\br
\end{tabular}
\end{indented}
\end{table}
The free energy of MnWO$_4$ corresponding to the Hamiltonian~(\ref{eq:Hamiltonian}) and obtained using mean field approach was thoroughly studied using Fourier transform and was found to give a correct phase transition sequence and good correspondence with the experimental data~\cite{Matityahu_MWO}. Furthermore, both constants sets of \tref{tab:Exchange_Constants_MnWO4} are shown to give results in qualitative agreement with the experimental data on MnWO$_4$~\cite{Matityahu_MWO}. For our Monte Carlo simulations below we use the exchange constants of~\cite{Ye_MWO} with $D_x=D$, $D_y=0$, and $D_z=-D$.

The Monte Carlo studies were performed using the Metropolis algorithm. At each temperature the system was allowed to relax for 5$\cdot10^9$ Monte Carlo steps, after which the data was averaged for 50$\cdot10^6$ steps. This approximately amounted to 8.7$\cdot10^5$ and 8.7$\cdot10^3$ steps per spin, respectively.

The phase transition sequence and the appearing magnetically ordered states in MnWO$_4$ can be described by magnetic order parameters belonging to the wave vector $\vec{k}_{\rm c}$~\cite{Ter-Oganessian_MnWO4}. Furthermore it was argued that the praphase approach to the description of phase transitions in MnWO$_4$ has significant advantages~\cite{Ter-Oganessian_Interpretation}. Thus in the following we use magnetic order parameters belonging to the wave vector $\vec{k}_{\rm c}$ of the orthorhombic praphase with the space group Pmcm (D$_{2h}^5$)~\cite{Ter-Oganessian_Interpretation}. The hypothetical phase transition Pmcm -- P2/c is described by the component $U_{XZ}$ of the homogenous deformation tensor, which has to be assumed nonzero value for the description of the monoclinic phase. In the following the axes $X$, $Y$, and $Z$ are assumed to be directed along the $a$ axis, the $b$ axis, and perpendicular to the $a$ and the $b$ axes, respectively. Choosing the praphase as the starting point for determination of the magnetic order parameters has no particular influence on the obtained results, but makes the present study consistent with our previous work on MnWO$_4$.

In the orthorhombic praphase structure MnWO$_4$ possesses two Mn$^{2+}$ ions in the unit cell located at $(\frac{1}{2},0.6853,\frac{1}{4})$ and $(\frac{1}{2},0.3147,\frac{3}{4})$, whose spins are denoted by $\vec{S}_1$ and $\vec{S}_2$, respectively. The magnetic representation in $\vec{k}_{\rm c}$ point of the Brillouin zone is given by 3P$_1$, where P$_1$ is a four-dimensional irreducible representation~\cite{Ter-Oganessian_Interpretation}. Upon the hypothetical phase transition Pmcm -- P2/c $U_{XZ}$ splits P$_1$ into two irreducible representations, which are denoted by G$_1$ and G$_2$ in the monoclinic structure. Denoting by $(g_{1\alpha},g_{2\alpha},g_{3\alpha},g_{4\alpha})$ ($\alpha=x,y,z$) the magnetic order parameters transforming according to P$_1$ and describing the spin components along the respective directions we obtain the manganese spins in a unit cell given by $n_1\vec{a}_1+n_2\vec{a}_2+n_3\vec{a}_3$ as
\begin{eqnarray}
\eqalign{
S_{1x}=& -\xi\left[(g_{1x}+g_{2x}-g_{3x}-g_{4x})\cos\left(\frac{n_1\pi}{2}\right)\right.\\
 &\qquad\left.-(g_{1x}-g_{2x}-g_{3x}+g_{4x})\sin\left(\frac{n_1\pi}{2}\right)\right], \\
S_{1y}=& -\xi\left[(g_{1y}-g_{2y}-g_{3y}+g_{4y})\cos\left(\frac{n_1\pi}{2}\right)\right.\\
 &\qquad\left.+(g_{1y}+g_{2y}-g_{3y}-g_{4y})\sin\left(\frac{n_1\pi}{2}\right)\right], \\
S_{1z}=&  \xi\left[(g_{1z}+g_{2z}+g_{3z}+g_{4z})\cos\left(\frac{n_1\pi}{2}\right)\right.\\
 &\qquad\left.-(g_{1z}-g_{2z}+g_{3z}-g_{4z})\sin\left(\frac{n_1\pi}{2}\right)\right], \\
S_{2x}=& -\xi\left[(g_{1x}-g_{2x}+g_{3x}-g_{4x})\cos\left(\frac{n_1\pi}{2}\right)\right.\\
 &\qquad\left.+(g_{1x}+g_{2x}+g_{3x}+g_{4x})\sin\left(\frac{n_1\pi}{2}\right)\right], \\
S_{2y}=& -\xi\left[(g_{1y}+g_{2y}+g_{3y}+g_{4y})\cos\left(\frac{n_1\pi}{2}\right)\right.\\
 &\qquad\left.-(g_{1y}-g_{2y}+g_{3y}-g_{4y})\sin\left(\frac{n_1\pi}{2}\right)\right], \\
S_{2z}=&  \xi\left[(g_{1z}-g_{2z}-g_{3z}+g_{4z})\cos\left(\frac{n_1\pi}{2}\right)\right.\\
 &\qquad\left.+(g_{1z}+g_{2z}-g_{3z}-g_{4z})\sin\left(\frac{n_1\pi}{2}\right)\right],
}\label{eq:MagneticRepresentation}
\end{eqnarray}
where $\vec{a}_1$, $\vec{a}_2$, and $\vec{a}_3$ are Bravais translations and $\xi=(-1)^{n2+n3}$.

Previous analysis of the exchange symmetry in MnWO$_4$ showed that the set of order parameters $(g_{1\alpha},g_{2\alpha},g_{3\alpha},g_{4\alpha})$ splits into two exchange multiplets $(g_{3x},g_{4x},g_{1y},g_{2y},g_{3z},g_{4z})$ and $(g_{1x},g_{2x},g_{3y},g_{4y},g_{1z},g_{2z})$~\cite{Ter-Oganessian_Interpretation}. In MnWO$_4$ the magnetic structures without applied magnetic field belong to the first exchange multiplet~\cite{Lautenschlager_MnWO4,Ter-Oganessian_Interpretation}. Further splitting within exchange multiplets consistent with the crystallographic symmetry is due to single-ion anisotropy $D_\alpha$.

\begin{figure}
\begin{indented}
\item[]\includegraphics[width=8cm]{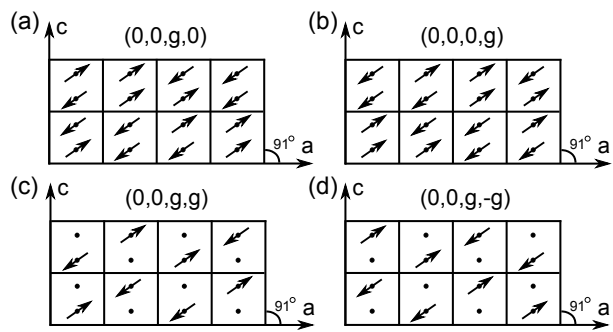}%
\end{indented}
\caption{Ordering patterns of Mn$^{2+}$ spins for various values of the order parameter $(g_{1y},g_{2y},g_{3x},g_{4x})$. The ordering pattern for negative value of $g$ is obtained from that for the positive value by spin inversion.\label{fig:0}}
\end{figure}

The theoretical model of phase transitions in MnWO$_4$~\cite{Ter-Oganessian_Interpretation}, which is in accordance with experimental data~\cite{Sagayama_MnWO4,Lautenschlager_MnWO4,Mitamura_MWO}, reveals that the magnetically ordered phases \textbf{AF3} and \textbf{AF2} are described by spatially modulated order parameters $(g_{3x}(\vec{r}),g_{4x}(\vec{r}))$ and $(g_{1y}(\vec{r}),g_{2y}(\vec{r}),g_{3x}(\vec{r}),g_{4x}(\vec{r}))$, respectively, whereas the commensurate phase \textbf{AF1} by the order parameter $(g_{3x},g_{4x})=(g,0)$. Therefore, the phase \textbf{AF1} can be described by four domains with $(g_{3x},g_{4x})$ equal to $(\pm g,0)$ and $(0,\pm g)$, which are shown in Figs.~\ref{fig:0}(a) and~(b). In this work we study the domain walls that separate these domains and are oriented perpendicular to the crystal axes $a$ and $c$, which we hereafter refer to as $a$-walls and $c$-walls, respectively. From the symmetry of the problem there are essentially three kinds of walls, those that separate the domains $(g,0)$ and $(0,g)$, the domains $(g,0)$ and $(-g,0)$, and the domains $(g,0)$ and $(0,-g)$, which we hereafter denote type~I,~II, and~III domain walls, respectively.

\begin{figure*}
\begin{indented}
\item[]\fl\includegraphics[width=15.5cm]{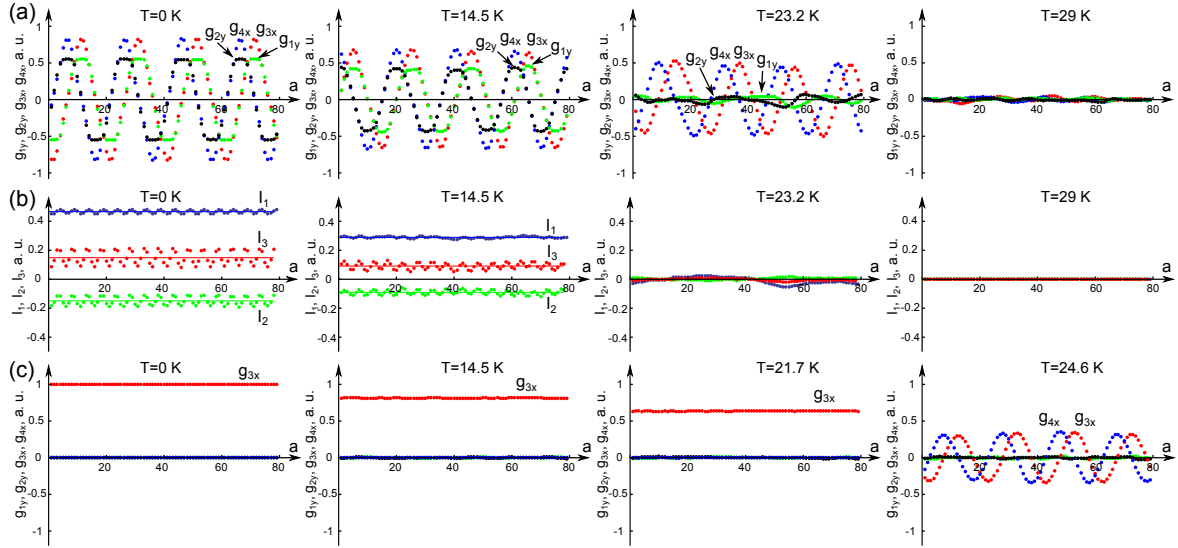}%
\end{indented}
\caption{(a) Spatial dependence of $(g_{1y},g_{2y},g_{3x},g_{4x})$ along the $a$ axis at decreasing temperature run. (b) Spatial dependence of $I_1$, $I_2$, and $I_3$ at different temperatures corresponding to Figure~(a). (c) Spatial dependence of $(g_{1y},g_{2y},g_{3x},g_{4x})$ along the $a$ axis during increasing temperature run, when \textbf{AF1} is chosen as the initial state. The horizontal axes in all figures are in cell units.\label{fig:1}}
\end{figure*}

Firstly, however, we study the sequence of magnetic phase transitions in MnWO$_4$ and for simplicity consider only the spatial variation of magnetic order along the $a$ axis by performing Monte Carlo simulations of a system with dimensions $80\times6\times6$ unit cells and periodic boundary conditions. \Fref{fig:1}(a) shows spatial variation of the order parameter $(g_{1y},g_{2y},g_{3x},g_{4x})$ in a decreasing temperature run. It can be seen that upon decreasing temperature a magnetic structure with spatially modulated order parameter $(0,0,g_{3x}(X),g_{4x}(X))$ appears at $T_N$. We associate this phase with \textbf{AF3}. The modulation period is approximately 20 unit cells giving the wave vector component of 0.2 along $a$, which roughly coincides with the experimental value 0.214. Simulations of larger systems ($1000\times6\times6$) gave approximately the same modulation period. The phase \textbf{AF3} is followed by the appearance of the $y$ component of spins $(g_{1y}(X),g_{2y}(X),g_{3x}(X),g_{4x}(X))$ and the resulting structure is associated with the phase \textbf{AF2}. Further temperature decrease does not result in a phase transition to a commensurate phase. We attribute the absence of such transition in our Monte Carlo simulations to its strongly first order character and suggest that further studies of this phase transition are needed, which is not in the scope of the present paper. However the commensurate state \textbf{AF1} is stable when chosen as an initial state. The respective simulations of the commensurate state during increasing temperature run are shown in \fref{fig:1}(c). It can be seen that the value of the nonzero component of the order parameter $g_{3x}$ gradually decreases with increasing temperature and at temperature between approximately 21.7~K and 24.6~K a first order phase transition to an incommensurate phase occurs. According to our simulations at $T=0$~K the internal energy of the commensurate state is lower than that of the modulated state, which ensures a phase transition to the commensurate state at temperature $T_1$ intermediate between $T=0$~K and $T_2$.

The macroscopic symmetry allows magnetoelectric interactions responsible for the emergence of electric polarization along the crystal $b$ axis, the lowest of which in powers of magnetic order parameters and spatial derivatives can be written in the form
\begin{eqnarray}
P_b&(g_{1x}g_{3y}+g_{2x}g_{4y}+g_{3x}g_{1y}+g_{4x}g_{2y}),\label{eq:ME}\\
P_b&\left(g_{1x}\frac{\partial g_{4y}}{\partial X} -  g_{2x}\frac{\partial g_{3y}}{\partial X} + g_{3x}\frac{\partial g_{2y}}{\partial X} - g_{4x}\frac{\partial g_{1y}}{\partial X} \right),\label{eq:InvDer1}\\
P_b&U_{XZ}\left(g_{1x}\frac{\partial g_{4y}}{\partial Z} -  g_{2x}\frac{\partial g_{3y}}{\partial Z} + g_{3x}\frac{\partial g_{2y}}{\partial Z} - g_{4x}\frac{\partial g_{1y}}{\partial Z} \right),\\
P_b&\left(-g_{1x}\frac{\partial g_{4y}}{\partial Z} +  g_{2x}\frac{\partial g_{3y}}{\partial Z} + g_{3x}\frac{\partial g_{2y}}{\partial Z} - g_{4x}\frac{\partial g_{1y}}{\partial Z} \right),\\
P_b&U_{XZ}\left(-g_{1x}\frac{\partial g_{4y}}{\partial X} +  g_{2x}\frac{\partial g_{3y}}{\partial X} + g_{3x}\frac{\partial g_{2y}}{\partial X} - g_{4x}\frac{\partial g_{1y}}{\partial X} \right).\label{eq:InvDer4}
\end{eqnarray}
Invariants similar to~(\ref{eq:InvDer1}) --~(\ref{eq:InvDer4}) exist, in which every term in the parantheses of the form $g_{ix}\partial g_{jy}/\partial\alpha$ should be substituted by $g_{jy}\partial g_{ix}/\partial\alpha$, where $i,j=1,2,3,4$ and $\alpha=X,Z$. According to our simulations and experimental data the relevant order parameter in MnWO$_4$ is $(g_{1y},g_{2y},g_{3x},g_{4x})$, with all other components of the order parameters $\{g_{i\gamma}\}$ ($i=1,2,3,4$, $\gamma=x,y,z$) experiencing only statistical fluctuations. Therefore, the magnetoelectric invariants~(\ref{eq:ME}) --~(\ref{eq:InvDer4}) can be written in simpler forms, which give  electric polarization $P_b$ proportional to the following combinations of order parameters
\begin{eqnarray}
P_b&\sim (g_{3x}g_{1y}+g_{4x}g_{2y})&=I_1,\\
P_b&\sim \left(g_{3x}\frac{\partial g_{2y}}{\partial\alpha} - g_{4x}\frac{\partial g_{1y}}{\partial\alpha} \right)&=I_2,\\
P_b&\sim \left(g_{2y}\frac{\partial g_{3x}}{\partial\alpha} - g_{1y}\frac{\partial g_{4x}}{\partial\alpha} \right)&=I_3,
\end{eqnarray}
where $\alpha=X,Z$. \Fref{fig:1}(b) gives $I_1$, $I_2$, and $I_3$ for several temperatures as functions of the cell number along the $a$ axis. The dependencies experience a sawtooth character and, therefore, the electric polarization should be proportional to average values, which are indicated by solid lines. Depending on the phase shift between $(g_{1y},g_{2y})$ and $(g_{3x},g_{4x})$, which is achieved during cooling of the system, the quantities $I_1$, $I_2$, and $I_3$ can have both positive and negative values giving both directions of electric polarization $P_b$.

\begin{figure}
\begin{indented}
\item[]\includegraphics[width=8cm]{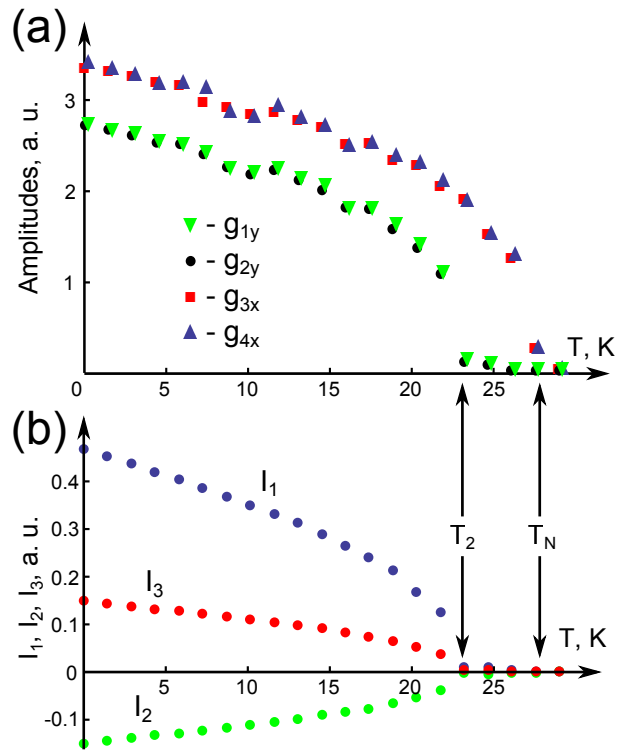}%
\end{indented}
\caption{(a) Temperature dependence of amplitudes of $g_{1y}$, $g_{2y}$, $g_{3x}$, and $g_{4x}$ in incommensurate states during decreasing temperature run. (b) Temperature dependence of the average values of $I_1$, $I_2$, and $I_3$ during decreasing temperature simulation.\label{fig:2}}
\end{figure}

\Fref{fig:2}(a) shows temperature dependencies of the amplitudes of modulated order parameter components $g_{1y}$, $g_{2y}$, $g_{3x}$, and $g_{4x}$ during decreasing temperature run. From these data one can estimate the phase transition temperatures as $T_N\approx26.7$~K and $T_2\approx23.2$~K. Our Monte Carlo studies give transition temperatures higher than those observed in experiments. We attribute this discrepancy to inaccuracy in exchange constants, which, however, from our point of view does not affect qualitative results obtained in the present work. \Fref{fig:2}(b) gives the temperature dependence of the average densities of $I_1$, $I_2$, and $I_3$, which are essentially proportional to the electric polarization. In agreement with experimental data $P_b$ appears at $T_2$ when all components of the order parameter $(g_{1y},g_{2y},g_{3x},g_{4x})$ condense.

\begin{figure*}
\begin{indented}
\item[]\fl\includegraphics[width=15.5cm]{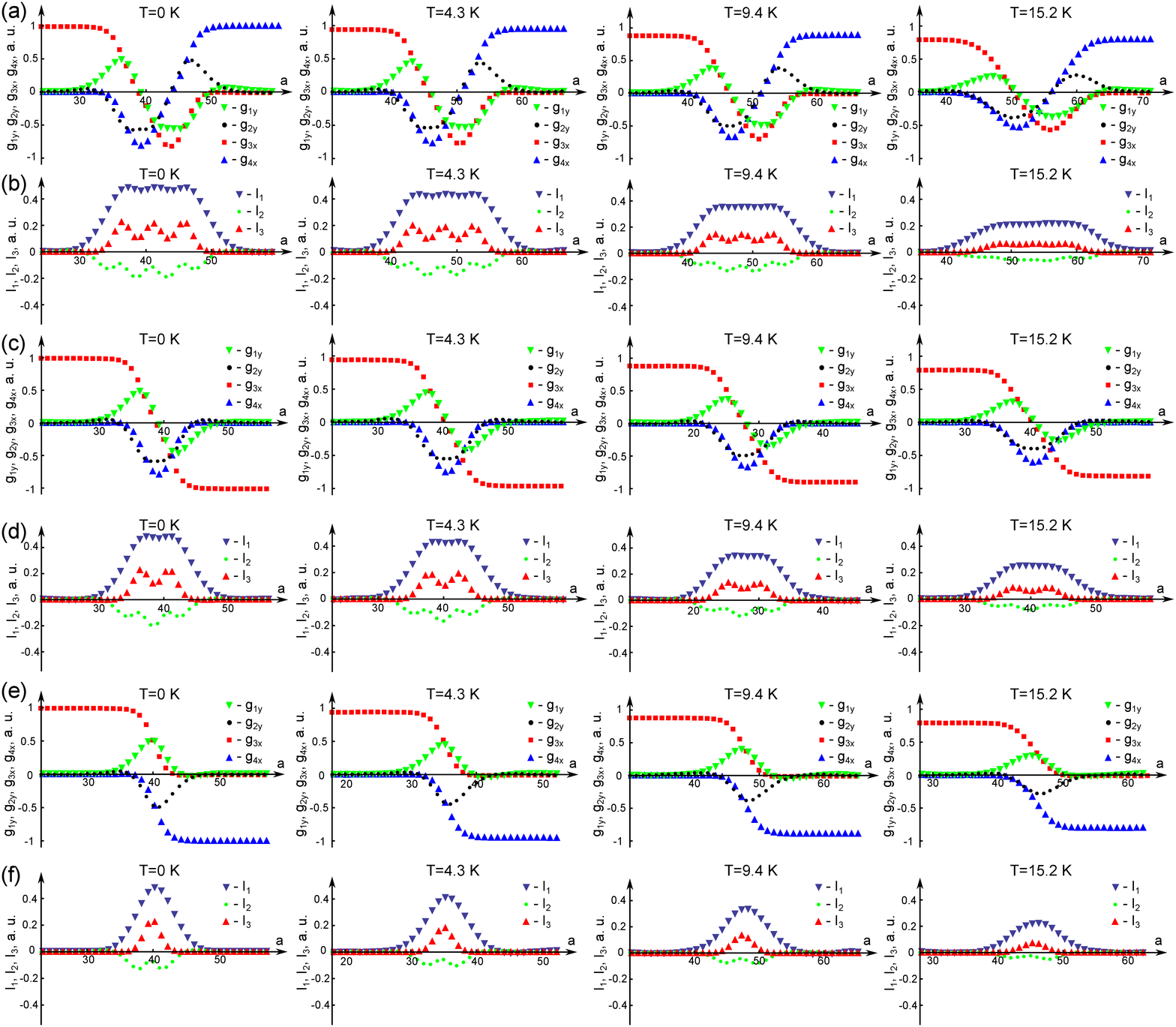}%
\end{indented}
\caption{(a) Spatial variation of $(g_{1y},g_{2y},g_{3x},g_{4x})$ across the type~I $a$-wall. (b) Spatial variation of $I_1$, $I_2$, and $I_3$ across the type~I $a$-wall. (c) and (d) are same as (a) and (b), but for the type~II wall, whereas (e) and (f) are for the type~III wall, respectively. The horizontal axes in all figures are in cell units.\label{fig:3}}
\end{figure*}

Next, we study the walls between the domains of the commensurate paraelectric phase \textbf{AF1}. \Fref{fig:3} shows variations of the order parameter and polarization across the domain walls of types~I,~II, and~III, which are oriented perpendicular to the $a$ axis of the crystal. For this purpose a system with dimensions $80\times6\times6$ unit cells was used. It can be seen that all domain wall types are ferroelectric. The values of $I_1$, $I_2$, and $I_3$ inside the walls are of the order of those in the incommensurate phase \textbf{AF2} shown in \fref{fig:1}(b), which allows concluding that the value of polarization of the walls should coincide with that of the bulk value in the phase \textbf{AF2}. Similar to the case above, depending on the signs of $g_{1y}$ and $g_{2y}$ the quantities $I_1$, $I_2$, and $I_3$ can have both positive and negative values, which can result in both directions of electric polarization $P_b$ of the wall. The states with positive and negative $P_b$ have the same energy.

\begin{figure*}
\begin{indented}
\item[]\fl\includegraphics[width=15.5cm]{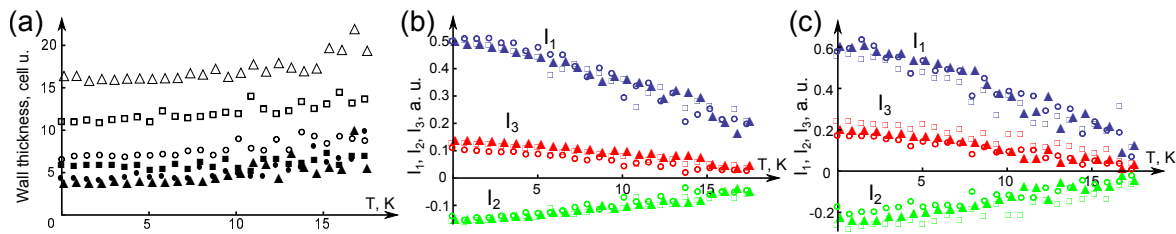}%
\end{indented}
\caption{(a) Temperature dependence of domain wall thickness. Empty and filled signs show data for domain walls oriented perpendicular to the crystal axes $a$ and $c$, whereas triangles, squares, and circles represent domain wall types~I,~II, and~III, respectively. (b) and (c) give temperature dependence of average values of $I_1$, $I_2$, and $I_3$ for domain walls oriented perpendicular to the crystal axes $a$ and $c$, respectively, whereas filled triangles, empty squares, and empty circles denote data for domain wall types~I,~II, and~III, respectively. \label{fig:4}}
\end{figure*}

The order parameter and electric polarization profiles across domain walls oriented perpendicular to the crystal $c$ axis are qualitatively similar to those of \fref{fig:3} and will not be given here.
\Fref{fig:4}(a) shows the temperature dependence of the domain wall thicknesses for both wall orientations, which we calculate as full width at half maximum of electric polarization. It can be found that the thickness of the $a$-walls, which is of the order of 7 -- 15 unit cells, is larger than than that of the $c$-walls, for which it takes values of the order of 5 unit cells. With increasing temperature the thickness of the walls gradually increases, whereas average polarization, which is shown in figures~\ref{fig:4}(b) and~(c), gradually decreases. The temperature evolution of the properties of domain walls in \fref{fig:4} are shown for temperatures up to approximately 17~K, because at higher temperatures the walls become less discernible and a phase transition to an incommensurate state occurs.

\section{Discussion\label{sec:Discussion}}

\begin{figure}
\begin{indented}
\item[]\includegraphics[width=8.0cm]{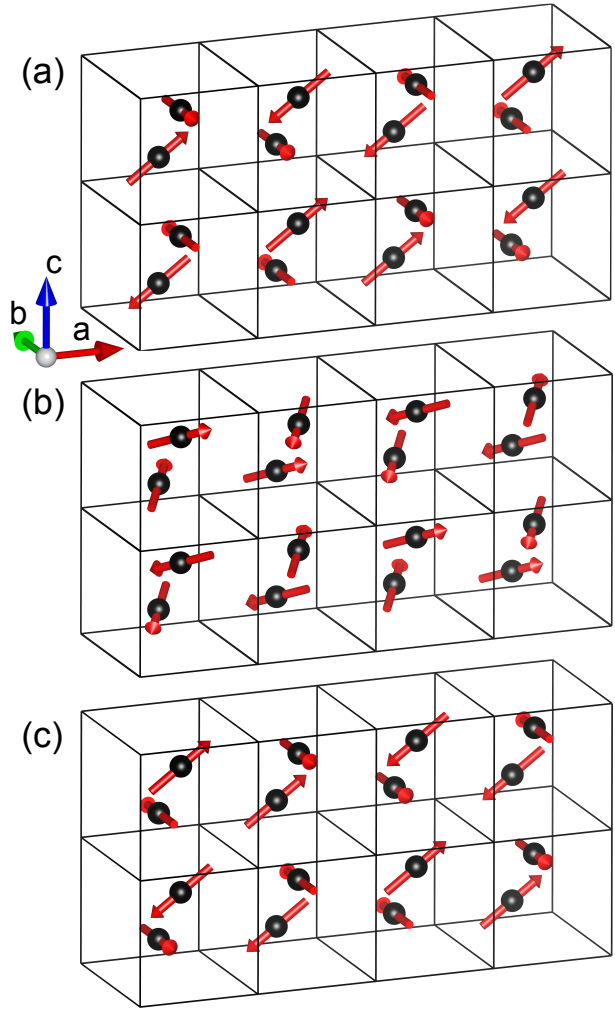}%
\end{indented}
\caption{Ordering patterns of Mn$^{2+}$ spins in the middle of the type~I~(a), type~II~(b), and type~III~(c) walls. The figure was created using the VESTA software~\cite{VESTA}.\label{fig:5}}
\end{figure}

In the previous section we have shown that the walls between the domains of the paraelectric phase \textbf{AF1} of MnWO$_4$ are ferroelectric with electric polarization along the $b$ axis of the crystal, which coincides with its direction in the phase \textbf{AF2}. It is known from the previous models of phase transitions in MnWO$_4$ that $P_b$ can only appear when both parts $(g_{1y},g_{2y})$ and $(g_{3x},g_{4x})$ of the order parameter $(g_{1y},g_{2y},g_{3x},g_{4x})$ condense, which follows from the form of magnetoelectric interactions~(\ref{eq:ME}) --~(\ref{eq:InvDer4})~\cite{Ter-Oganessian_Interpretation,Toledano_MnWO4}. In our simulations we find that inside the walls both parts of the order parameter condense in such a way that the wall becomes polar. The appearance of nonzero $(g_{1y},g_{2y})$ inside the wall can be explained in the following simple way. Consider, for example the type~I wall. The change from one domain with the order parameter $(0,0,g,0)$ to the other with $(0,0,0,g)$ without nonzero $(g_{1y},g_{2y})$ would require passing through a state with the order parameter $(0,0,g',g')$. Such state, however, results in ordering of only one of the two spins $\vec{S}_1$ and $\vec{S}_2$ in every unit crystal cell as follows from~(\ref{eq:MagneticRepresentation}) and is shown in Fig.~\ref{fig:0}(c). (Figure~\ref{fig:0}(d) shows the phase state $(0,0,g',-g')$.) Therefore, this state possesses considerably higher exchange energy and is less favorable than the state with nonzero $(g_{1y},g_{2y})$ provided that the single-ion anisotropy is sufficiently small, which is the case in MnWO$_4$. In terms of irreducible representations of the monoclinic space group the phase \textbf{AF1} is described by G$_2$, whereas G$_1$ additionally appears inside the walls between the domains of the phase \textbf{AF1}. Simultaneous appearance of G$_1$ and G$_2$ inside the wall results in its ferroelectric polarization~\cite{Ter-Oganessian_Interpretation}. The ordering patterns of Mn$^{2+}$ spins in the middle of the type~I,~II, and~III walls, which approximately correspond to the phase states $(-g,-g,-g,-g)$, $(0,-g,0,-g)$, and $(g,-g,g,-g)$, respectively, are shown in Fig.~\ref{fig:5}.

We have also identified three possible macroscopic magnetoelectric interactions, which give electric polarization proportional to $I_1$, $I_2$, and $I_3$. It follows from our Monte Carlo studies that all three possible magnetoelectric interactions can contribute to ferroelectric polarization in the phase \textbf{AF2} as well as in  domain walls. However, it could be the case that one or two of them prevail over the others. Besides, these interactions give different results for commensurate states, i.e. the interactions possessing spatial derivatives ($I_2$ and $I_3$) cannot give electric polarization in commensurate magnetically ordered state, whereas $I_1$ can induce polarization in commensurate as well as incommensurate magnetic states. Therefore, it is essential to determine which of these interactions are responsible for magnetoelectric properties of MnWO$_4$. Here we argue that experimental studies of the walls between the domains of the phase \textbf{AF1} can clarify this problem. Indeed, the crystal structure of MnWO$_4$ possesses ...-Mn-O-Mn-O-... zigzag chains running along the $c$ axis, whereas along the $a$ and $b$ axes the interaction between manganese spins is more complex (Mn-O-W-O-Mn). Thus, the exchange paths along the $a$ and the $c$ axes are substantially different. Therefore, one can argue that if the magnetically induced electric polarization in MnWO$_4$ is due to differences of the order parameter in neighboring cells, i.e. if the relevant magnetoelectric interactions in MnWO$_4$ are proportional to the spatial derivatives of the order parameter as in $I_2$ and $I_3$, then the $a$-walls should possess zero or low electric polarization in contrast to the $c$-walls, which should be ferroelectric with electric polarization comparable to that in the bulk. On the contrary, if similar polarization is revealed experimentally for both the $a$- and the $c$-walls, then one can argue that the relevant magnetoelectric interaction is~(\ref{eq:ME}).

In the available literature the studies of the domains or domain walls in MnWO$_4$ are limited to those in the ferroelectric phase \textbf{AF2} and are mostly concerned with the dynamics of electric polarization switching by electric or magnetic fields~\cite{Hoffmann_MnWO4,Taniguchi_MnWO4,Niermann_MnWO4}. In~\cite{Yu_MnWO4_ElectricField} the effect of poling electric field on ferroelectric properties of MnWO$_4$ was studied. The authors observed nonzero electric polarization below $T_1$ when the sample was cooled down from $T\gg T_N$ under poling electric field of 10~kV~cm$^{-1}$ to temperatures below $T_1$, after which the field was removed and a residual polarization was observed. This phenomenon was interpreted as enhancement of stability of the phase \textbf{AF2} by electric field, which results in surviving of the clusters of ferroelectric phase \textbf{AF2} in the paraelectric phase \textbf{AF1}. The authors estimated the volume fraction of the phase \textbf{AF2} at $T=4$~K to be as high as 13\% by comparing the polarization value to that of the pure phase \textbf{AF2}. We agree with the authors of~\cite{Yu_MnWO4_ElectricField} that electric field enhances the stability of the phase \textbf{AF2}. This enhancement, however, is present only when the poling field is applied to the sample and should disappear upon its removal, which at $T<T_1$ should result in a phase transition of the remains of the phase \textbf{AF2} to the phase \textbf{AF1} and disappearance of polarization. From our point of view the presence of electric polarization at temperatures below $T_1$ can be interpreted as polarization of walls between the domains of the phase \textbf{AF1}. The poling electric field present upon the phase transition to paraelectric phase \textbf{AF1} from \textbf{AF2} should (i) pole the appearing domain walls and align their polarization along the field, and (ii) should in general increase the number of domains and, respectively, the number of the walls between them. Therefore, we argue that these experimental results can be interpreted as supporting the picture of polar domain walls in the phase \textbf{AF1} suggested in the present work.

The analogy in the phase transition sequences of MnWO$_4$ and CuO, which is also reflected in similarity of their macroscopic descriptions~\cite{Ter-Oganessian_Interpretation,Quirion_MWO_CuO}, argues that the same phenomena suggested for the low-temperature phase of MnWO$_4$ in the present work can take place in the phase \textbf{AF1} of CuO. Electric polarization of domain walls in the phase \textbf{AF1} of CuO will extend the exploitable temperature range of magnetoelectric properties of CuO down to 0~K from the rather narrow range 213 -- 229~K of stability of the phase \textbf{AF2}.

\section{Conclusions\label{sec:Conclusions}}

We have shown that the walls between the domains of the low-temperature commensurate paraelectric phase \textbf{AF1} of MnWO$_4$ and CuO are ferroelectric. The value and direction of electric polarization inside the walls coincide with those of the magnetoelectric phase \textbf{AF2} of the respective compounds. Experimental observation of electric polarization of domain walls in MnWO$_4$ should help clarifying the macroscopic interaction responsible for its magnetoelectric properties: flexomagnetoelectric or trilinear. Electric polarization of domain walls of CuO significantly extends the temperature range of its exploitable magnetoelectric properties. Together with the possibility of switching the electric polarization of the walls and the complex behavior of these magnetoelectrics in magnetic fields this may open new interesting phenomena in the physics of these magnetoelectrics.

\ack
The authors acknowledge the financial support from the SFedU Grant no. 213.01-2014/011-VG and RFBR Grant no. 12-02-31229-mol\_a.

\section*{References}


\end{document}